\documentclass[doublecol]{epl2}
\usepackage{amssymb,amsmath,latexsym}
\usepackage{epsfig}
\def\Fbox#1{\vskip1ex\hbox to 8.5cm{\hfil\fboxsep0.3cm\fbox{%
  \parbox{8.0cm}{#1}}\hfil}\vskip1ex\noindent}  %%  {TEXT} in BOX

\DeclareMathAlphabet{\mathitbf}{OML}{cmm}{b}{it}

\newcommand{\rv}{\mathitbf r}
\newcommand{\Sv}{\mathitbf S}

\newcommand{\uv}{\mathitbf u}

\newcommand{\vv}{\mathitbf v^{(\gamma, r)}}
\newcommand{\vs}{\mathitbf v^{(\gamma, \phi )}}
\newcommand{\vvb}{\mathitbf v^{(b,r)}}
\newcommand{\vsb}{\mathitbf v^{(b,\phi )}}

\newcommand{\calBold}[1]{\mbox{\boldmath${\cal #1}$}}
\newcommand{\mathBold}[1]{\mbox{\boldmath$#1$}}
\newcommand{\Xv}{\mathBold{\Xi}}
\newcommand{\Hes}{\calBold{H}}
\newcommand{\Trrr}{\calBold{T}^{(rrr)}}
\newcommand{\Trrs}{\calBold{T}^{(rr\phi )}}
\newcommand{\Trsr}{\calBold{T}^{(r\phi r)}}
\newcommand{\Tsrr}{\calBold{T}^{(\phi rr)}}
\newcommand{\Tsss}{\calBold{T}^{(\phi \phi \phi )}}
\newcommand{\Tssr}{\calBold{T}^{(\phi \phi r)}}
\newcommand{\Tsrs}{\calBold{T}^{(\phi r\phi )}}
\newcommand{\Trss}{\calBold{T}^{(r\phi \phi )}}
\newcommand{\Tesr}{\calBold{T}^{(r)}}
\newcommand{\Tess}{\calBold{T}^{(\phi )}}

\newcommand{\eigK}[1]{\calBold{\psi}^{(#1)}}
\newcommand{\B}[1]{{\bm{#1}}}
\let \= \equiv \let\*\cdot \let\~\widetilde \let\^\widehat \let\-\overline

\begin{document}
\title{The Plastic Response of Magnetoelastic Amorphous Solids}
\author{H.G.E.Hentschel$^{1,2}$, Valery Ilyin$^1$ and Itamar Procaccia$^1$}
\institute{$^1$Dept. of Chemical Physics, The Weizmann Institute of
Science, Rehovot 76100, Israel
\\$^2$Dept of Physics,
Emory University, Atlanta, Georgia, }

\abstract{
We address the cross effects between mechanical strains and magnetic fields on the plastic response of magnetoelastic amorphous solids.
It is well known that
plasticity in non-magnetic amorphous solids under external strain $\gamma$ is dominated by the co-dimension 1 saddle-node bifurcation in which an eigenvalue of the Hessian matrix vanishes at $\gamma_P$ like
$\sqrt{\gamma_P-\gamma}$. This square-root singularity determines much of the statistical physics of elasto-plasticity, and in particular that of the stress-strain curves under athermal-quasistatic conditions. In this Letter we discuss the much richer physics that
can be expected in magnetic amorphous solids. Firstly,  magnetic amorphous solids exhibit co-dimension 2 plastic instabilities, when an external strain and an external magnetic field are applied simultaneously. Secondly, the phase diagrams promise a rich array of
 new effects that have been  barely studied; this opens up a novel and extremely rich research program for magnetoplastic materials.}
\maketitle
%%%%%%%%%%%%%%%%%%%%%%%%%%%%%%%%%%
{\bf Introduction}: The well known magnetostriction effect where switching on a magnetic field changes the volume of
a sample of magnetoelastic solid and the Villari effect in which mechanical strain changes the magnetization are just two examples of the rich variety of cross effects that exist in magnetoelastic materials when
both an external mechanical strains and a magnetic field are employed. Surprisingly, the fundamental physics of plasticity has been studied much more extensively in the context of pure mechanical strains in non-magnetic amorphous solids, with many experiments and numerical simulations becoming available in recent decades. A much richer physics of plasticity can, however, be expected in magnetoelastic amorphous materials like metallic glasses, and the aim of this letter is to commence a research program in this direction.

The fundamental physics of plastic instabilities in non-magnetic amorphous solids has been uncovered in recent years in the context
of athermal, quasistatic mechanical (AQS) strain. While many experiments are done at finite temperature and finite strain rates,
AQS studies afford a unique laboratory for exposing the clean fundamental physics of plastic instabilities. In particular
bifurcation theory, sometime known as catastrophe theory, provides a powerful
framework for the discussion of the singular behavior of such instabilities. For purely mechanical strains irreversible plastic events occur in AQS conditions when an eigenvalue of the hessian matrix $\Hes$ hits zero. The Hessian matrix is defined as
\begin{equation}
{\cal H}_{ij} \equiv \frac{\partial U}{\partial \rv_i\partial \rv_j}\ , \quad Nd\times Nd ~\text{  symmetric matrix}
\end{equation}
where $U(\{\rv_i(\gamma )\}, \gamma)$ is the total potential energy of a solid containing $N$ particles whose coordinates depend on the external strain $\gamma$, i.e. $\{\rv_1(\gamma)\dots\rv_N(\gamma)\}$. The generic plastic event \cite{98ML,04ML} is a codimension-1 saddle-node bifurcation which
occurs at some value of $\gamma=\gamma_P$ of the external strain where the lowest eigenvalue of the Hessian matrix
(excluding Goldstone modes when they exist) hits zero like $\sqrt{\gamma_P-\gamma}$. The saddle-node bifurcation has been shown to be universal for a large variety of amorphous solids from simple binary glasses with pair potentials to metallic glasses with many-body interaction potentials \cite{12DKP}. Among other interesting
phenomena this square-root singularity determines the system-size dependence of the average stress drops
$\langle \Delta \sigma \rangle$ and average energy drops $\langle \Delta U \rangle $ during the plastic events in the elasto-plastic steady state of amorphous solids above the yield stress \cite{10KLP},
\begin{equation}
\langle \Delta \sigma \rangle \sim N^\beta\ , \quad \langle \Delta U \rangle\sim N^\alpha \ ,
\end{equation}
with universal exponents $\alpha=1/3$, $\beta=-2/3$.

{\bf Modeling magneto-plasticity}: The aim of this Letter is to discuss the rich and interesting physics of plastic instability and
failure in general in
magnetic amorphous solids in which two control parameters exist, i.e. the external strain $\gamma$ and a magnetic field $B$. We cannot cover here all the exciting physics that involve cross-effects between strain $\gamma$ and magnetic field $B$ in phenomena such as magnetostriction, the Villari effect, Barkhausen noise, magnetic domain evolution and consequently their effects on the mechanical properties of materials like metallic glasses \cite{33STE,80KASK,82Liv,95Gil,99DZ,00LACIM}.

 Our aim in this Letter is to point out to the interested community that magnetoelastic materials have a lot to offer in terms of new riddles, and here we focus on the $B-\gamma$ phase diagrams including the existence of codimension-2 plastic instabilities. Possible totally new singularities will be discussed at the end.

In amorphous solids whose constituents are magnetic the potential energy $U$ consists of two parts
\begin{eqnarray}
\label{u}
&&U(\{\rv_i(\gamma, B )\}, \{{\Sv}_i(\gamma, B)\}, \gamma, B) = U_{\rm mech}(\{\rv_i(\gamma, B )\}, \gamma)\nonumber\\&&+U_{\rm mag}(\{\rv_i(\gamma, B )\},\{{\Sv}_i(\gamma, B)\}, \gamma, B),
\end{eqnarray}
where ${\Sv}_i$ are spin variables and the mechanical part can be written as
\begin{equation}
\label{umech}
U_{\rm mech}(\{\rv_i(\gamma, B )\}, \gamma) = \sum_{<ij>} u(r_{ij}(\gamma, B) ,\gamma) \ ,
\end{equation}
where $r_{ij}\equiv |\rv_i-\rv_j|$. (in this Letter we shall treat both the strain $\gamma$ and magnetic field B as scalars). The
interparticle potential typically exhibits a minimum at a scale $r_{ij}=\sigma$. Note that most glassy materials are made of
a number of constituents and therefore this scale will be different for each distinct pair of constituents.

The magnetic part needs to be modeled to fit best a particular material, and different materials will have somewhat different magnetic interactions. For concreteness therefore let us consider here
an amorphous 2-dimensional model of classical $xy$ spins in which the orientation of each spin is given by an angle $\phi_i$. We also denote by $\theta_i({\bf r}_i)$ the local preferred easy axis of anisotropy, and end up with the magnetic contribution to the potential energy in the form \cite{footnote}:
\begin{widetext}
\begin{equation}
\label{potential}
U_{\rm mag}(\{{\rv}_i\}, \{{\Sv}_i\}) = - \sum_{<ij>}J(r_{ij}) \cos{(\phi_i-\phi_j)}- K \sum_i \cos{(\phi_i-\theta_i(\{{\rv}_i\}))}^2-  B \sum_i \mu_i \cos{(\phi_i)} \ .
\end{equation}
\end{widetext}
where $J(r_{ij})$ is the exchange interaction overlap integral whose $r_{ij}$ dependence needs to be found for each material. As shown in appendix A, the exchange integral is expected to have a relatively sharp peak at a typical scale $r^*$. Of crucial importance is the
relative dimensionless ratio $r^*/\sigma$ where $\sigma$ is the position of the minimum set by interparticle potential (in a typical
glass this ratio will be different for every given pair of constituents). Consider for example a small uniaxial compressive stress on
a given material. For $r^*/\sigma>1$ this will result in decreasing the magnetic interactions with increase of compressive stress. On the other hand for $r^*/\sigma<1$ compressive stress will result in increasing the magnetic interactions. Indeed, in this case it is even possible that larger compressive stress will turn around this behavior. In the case of tensile stresses the oppositive behavior can
be expected; while for shear stresses anisotropic phenomena might appear.

The second term of Eq.~(\ref{potential})
that models the effect of local anisotropy will tend to form magnetic domains, depending on the relative magnitude of the parameter $K$ with respect to $Jq$ where $q$ is the average number of nearest-neighbors (note that $K$ could in principle also be a function of $\rv_{ij}$, but we expect that the existence of local anisotropy due to the local positions of the atomic neighborhood is more important.) For $K/Jq>1$ we expect a complex interplay between domain formation, a vortex glass phase and ferromagnetic behavior.

The last term is
the interaction with the external field $B$. here $\mu_i \equiv\mu_B g_i L_i$ where $\mu_B$ is the
Bohr magneton, $L_i$ the angular momentum of the state considered and $g_i$ is the associated Lande g-factor. Note that the first two terms will exhibit important coupling between strain and magnetism due to the dependence on $\rv_{ij}$. Note also that the macroscopic magnetization $\B M$ of our sample is $\B M=\frac{1}{V}\sum_i {\B \mu}_i$.

At this point the joint dependence of the potential $U$ on the positions $\rv_i$ and the angles $\phi_i$ clarifies; it implies that strain $\gamma$ and magnetic field $B$ can be changed quasi-statically to maintain the magnetic amorphous solid in mechanical equilibrium. This requires that the net forces and torques on each particle vanish ,
\begin{equation}
\label{constraint}
\frac{\partial U}{\partial \rv_i}  =  0 \ , \quad
\frac{\partial U}{\partial \phi_i}  =  0  ,
\end{equation}
for all strains and magnetic fields. Thus, if the system is subject to an affine transformation
$ \rv_i \rightarrow \mathBold{h}(\gamma)\cdot\rv_i$, the particles will exhibit additional non-affine displacements to annul the resulting forces. but then also the spin orientations will change to minimize the energy, and in total we expect a non affine response according to
\begin{equation}
\label{ustrain}
\rv_i   \to  \mathBold{h}(\gamma )\cdot\rv_i + \uv_{i,NA} \ , \quad
\phi_i  \to  \phi_i + \phi_{i,NA}\ .
\end{equation}
Similarly, in response to a change in the applied magnetic field $B$ both the particles positions and the spin orientation will change; we denote the total changes that are required to keep the system in mechanical equilibrium as $\uv_{i,NB}$ and $\phi_{i,NB}$:
\begin{equation}
\label{umag}
\rv_i   \to  \rv_i + \uv_{i,NB} \ , \quad
\phi_i  \to  \phi_i + \phi_{i,NB}\ .
\end{equation}

{\bf Plastic instabilities}:
To proceed further, we need to study the generalized Hessian matrix $\Hes$, which is now of rank $N(d+1)\times N(d+1)$:
\begin{equation}
\label{Hesa}
\Hes =
\begin{pmatrix}
  \Hes^{(rr)} & \Hes^{(r\phi )}\\
  \Hes^{(\phi r)} & \Hes^{(\phi \phi )}
\end{pmatrix}
\end{equation}
Here the four partial matrices are defined by
\begin{eqnarray}
\label{hessian}
\Hes^{(rr)}_{ij} & \equiv  & \frac{\partial^2U}{\partial \rv_i\partial \rv_j}  \qquad  \text{($dN \times dN$ matrix)}\nonumber \\
\Hes^{(r\phi )}_{ij} & \equiv  & \frac{\partial^2U}{\partial \rv_i\partial \phi_j} \qquad  \text{($dN \times N$ matrix)}\nonumber \\
\Hes^{(\phi r)}_{ij} & \equiv  & \frac{\partial^2U}{\partial \phi_i\partial \rv_j} \qquad  \text{($N \times dN$ matrix)}\nonumber \\
\Hes^{(\phi \phi)}_{ij} & \equiv  & \frac{\partial^2U}{\partial \phi_i\partial \phi_j} \qquad  \text{($N \times N$ matrix)} \ .
\end{eqnarray}
The Hessian matrix $\Hes$ is real and symmetric, and therefore diagonalizable. Besides Goldstone modes with zero eigenvalues all the other modes are associated with positive eigenvalues as long as the
system is mechanically stable. Plastic instabilities will occur when the lowest positive eigenvalue
of $\Hes$ will approach zero upon the increase of $\gamma$, $B$, or both. These instabilities are manifest in the equations of motion of the non-affine responses defined in eqs.(\ref{ustrain}) and (\ref{umag}). These are denoted as
\begin{eqnarray}
\label{flows}
\vv_j   &=&  \frac{\partial {\bf u}_{j,NA}}{\partial \gamma} {\bf |}_{B} \ , \quad
\vs_j   =  \frac{\partial {\phi}_{j,NA}}{\partial \gamma} {\bf |}_{B}  \ , \nonumber\\
\vvb_j  & = & \frac{\partial {\bf u}_{j,NB}}{\partial B} {\bf |}_{\gamma}  \ , \quad
\vsb_j   =  \frac{\partial {\phi}_{j,NB}}{\partial B} {\bf |}_{\gamma} \ .
\end{eqnarray}
Note the physical meaning of these non-affine flows. The first is the well known non-affine response which occurs also in non-magnetic amorphous solids as a result of an external strain. There is now a new response to the external strain, namely that of the spins. Finally we have two new
non-affine responses to the magnetic field, i.e  the response of the spins and the response of the particle positions.  One can derive the equation of motion for these flows in the form
\begin{eqnarray}
\label{nonAffineVelocities}
{\bf v}^{(\gamma )} & = & -\Hes^{-1}\cdot \Xv^{(\gamma )} \nonumber \\
{\bf v}^{(b)} & = & -\Hes^{-1}\cdot \Xv^{(b)} .
\end{eqnarray}
where the mismatch forces are defined in Appendix B and
\begin{equation}
\label{va}
{\bf v}^{(\gamma )} =
\begin{pmatrix}
  \vv\\
  \vs
\end{pmatrix}
\ , \quad {\bf v}^{(b)} =
\begin{pmatrix}
  \vvb\\
  \vsb
\end{pmatrix}
\end{equation}
It is now manifest in Eq. (\ref{nonAffineVelocities}) that when an eigenvalue of $\Hes$ hits zero
the non-affine response can be very large indeed.

The derivation of the equations of motion of the eigenvalues of the Hessian matrix follows verbatim the
techniques uses in the purely mechanical case \cite{11HKLP}. These read
\begin{eqnarray}
\label{eigenvalueDerivative}
\frac{\partial \lambda_k}{\partial\gamma}{\bf |}_{B}   & =  & c^{(\gamma )}_{kk} - \sum_\ell \frac{a^{(\gamma )}_\ell [ b^{(r)}_{kk\ell}+b^{(\phi )}_{kk\ell}]}{\lambda_\ell}\nonumber \\
\frac{\partial \lambda_k}{\partial B}{\bf |}_{\gamma}  & =  & c^{(b)}_{kk} - \sum_\ell \frac{a^{(b)}_\ell [b^{(r)}_{kk\ell}+b^{(\phi )}_{kk\ell}]}{\lambda_\ell} .
\end{eqnarray}
The coefficients in these equations are all defined in Appendix B.

%%%%%%%%%%%%%%%%%%%%%%%%
{\bf Symmetries and Codimension-2 Instabilities}:
In the rest of this Letter we focus on the novel co-dimension 2 plastic instabilities that arise in
the present model. To this aim note that while the potential energy $U$ has no special behavior under $\gamma \rightarrow - \gamma$ for each configuration of $\{\rv_i \}, \{{\phi}_i \}$, it does remain invariant under the combined
transformations
\begin{equation}
\label{trans}
[\{\rv_i \}, \{{\phi}_i \};\gamma, B] \rightarrow [\{\rv_i \}, \{{\phi}_i + \pi \}; \gamma, -B].
\end{equation}
This symmetry has important implication for the properties of strained amorphous solids in the presence of magnetic fields. One immediate consequence is that the eigenvalues of the Hessian matrix obey this symmetry as well,
\begin{equation}
\label{lambda}
\lambda_k (\{\rv_i\},-\B M ;\gamma, -\B B) = \lambda_k (\{\rv_i\},\B M;\gamma, \B B)\ ,
\end{equation}
where we have now returned the vector notation to the magnetic field. Consider now the $\B B,\gamma$
parameter space, firstly along the line $B=0$. Along this line we expect to find a first
plastic instability at a value of $\gamma=\gamma_P$. It is easy to see from Eq. \ref{eigenvalueDerivative} that for $B=0$ and when one of the eigenvalues tends to zero, say $\lambda_P\to 0$, the solution has the
 square-root singularity, $\lambda_P \sim \sqrt{\gamma_P-\gamma}$, which is the hallmark of a saddle node bifurcation. From this point there emanates a line of instabilities at values of $\gamma=\gamma_P(B)$ where (again from examining Eq. (\ref{eigenvalueDerivative})),
\begin{equation}
\lambda_P(\gamma, B) \sim \sqrt{\gamma_P(B)-\gamma} \ . \label{gamp}
\end{equation}
Due to the symmetry (\ref{lambda}) we expect this line to depend on the value of the magnetic field according to
\begin{eqnarray}
\label{a}
\gamma_P(B)  & = & \gamma_P  - k_1 {\bf M\cdot B} - k_2 B^2+\dots \qquad \text {if }\B M\ne 0 \nonumber \\
\gamma_P(B)  & = & \gamma_P  - k_2 B^2 -k_4 B^4 +\dots\qquad \text {if }\B M=0  .
\end{eqnarray}
A consequence of Eq.~(\ref{a}) is that for a small magnetic field $B$ the lowest plastic mode will become unstable at a finite magnetic field
\begin{eqnarray}
\label{b}
B_P(\gamma )  & = & (\gamma_P-\gamma)/(k_1 M) \qquad \text {if }\B M\ne 0 \nonumber \\
B_P(\gamma )  & = & \sqrt{ (\gamma_P-\gamma)/k_2} \qquad \text {if }\B M=0 \  .
\end{eqnarray}
for $\gamma < \gamma_p$ but close to it. Inserting Eq. (\ref{b}) in Eq. (\ref{a}) and then in (\ref{gamp}), and then expanding to first order, we find
\begin{eqnarray}
\label{7}
\lambda_P(\gamma , B)  & \sim & \sqrt{M}\sqrt{(B_P(\gamma )-B)} \qquad \text {if }\B M\ne 0 \\
\lambda_P(\gamma , B)  & \sim & (\gamma_P-\gamma)^{1/4}\sqrt{(B_P(\gamma )-B)}  \qquad \text {if }\B M=0 \nonumber  \ .
\end{eqnarray}
At this point we examine a co-dimension 2 path in the $\gamma,B$ plane, denoted as $B(\gamma)$. Along
this path the eigenvalue satisfies the equation
\begin{equation}
\label{lgb2}
d\lambda_P/d\gamma =  (\frac{\partial \lambda_P}{\partial\gamma}{\bf |}_{B})  + (\frac{\partial \lambda_P}{\partial B}) {\bf |}_{\gamma} (dB(\gamma)/d\gamma ) .
\end{equation}
Now if the curve $B(\gamma )$ is analytic the saddle-node exponents will not change, but consider the singular curve in the $(\gamma , B)$ plane that ends at $(\gamma_P , B_P)$ of the form
\begin{equation}
\label{lgb3}
(B_P - B) = K (\gamma_P (B_P) - \gamma )^{\delta} .
\end{equation}
Now $ (dB/d\gamma ) = -\delta  K (\gamma_P (B_P) - \gamma )^{\delta -1}$ and different behavior can be expected for $\delta < 1$ and $\delta >= 1$. If $\delta >= 1$ the saddle-node exponents remain unchanged but if $\delta < 1$ then
\begin{equation}
\label{lgb4}
\lambda_P \sim (\gamma_P (B_P) - \gamma )^{\delta/2} .
\end{equation}
This is the main result of the present calculation, showing that co-dimension 2 paths can yield
a range of singularities to the lowest eigenvalue provided $\delta<1$. As a consequence the exponent characterizing
the approach of $\lambda_P$ to zero can now in principle take any value smaller than a 1/2.

The discussion so far has been based on two implicit assumptions that we want to discuss here further. The first assumption concerns the thermodynamic limit $N\to \infty$. Because we are dealing with amorphous solids we might expect that even in the case that $M=0$
in the thermodynamic limit, for finite sample $M\sim 1/\sqrt{N}$. Thus in the presence of small magnetic fields the first of Eqs. (18)
is always dominant. The other assumption was that the first plastic instability at $\gamma_P$ is well separated from a second plastic
instability at, say, $\bar\gamma_P$. When B increases the instability lines emanating from these two points at $B=0$ can come
closer and maybe even cross. This is definitely an interesting bifurcation where two eigenvalues go to zero simultaneously,
bringing about the need for a new theory and new consideration \cite{12HP}.

The existence of a range of scaling behavior for the eigenvalues of the Hessian that hit zero has
important implications on the statistical physics of magnetoplastic materials, putting them in a separate universality class from the strictly mechanical instabilities seen in elastoplastic materials. We note that we only discussed the case of weak magnetic fields where the nonlinear terms in Eq. (\ref{a}) are negligible. Note that in these equations the sign of the various coefficients $k_1, k_2$ and $k_4$ is crucially important in
determining different possible phenomenologies. Second order perturbation theory suggests that
$k_2>0$ \cite{12HP}.  Depending on the signs of these coefficient there may exist a point  $(\gamma_c, B_c)$ at which this line terminates \cite{12HP}. In such a case we expect to see very interesting
scaling behavior whose precise nature must await future work.

\section{Appendix A: The typical exchange integral}

For solids where only small deviations from the
equilibrium positions occur this integral can be given by the two first terms
of a Taylor expansion around the equilibrium atomic spacing $r_0$ at zero
temperature (see, e.g., \cite{CK86,GCGPL96,GR96})
\begin{equation}
J(r)=J_0+J^\prime(r-r_0),
\label{tp}
\end{equation}
where the constant $J^\prime$ is responsible for coupling between elastic
and magnetic energies and can be positive or negative.
For amorphous solids in which the particles may migrate relatively large distances this is not a sufficient model.  The simplest model that appears consistent with observation has
an exponential long-range decay (see, e.g., \cite{BHH00,CB79}) and a rapid rise at short distance, for example
\begin{eqnarray}
J(r)&=&J_0e^{-\beta (r-r_1)},\hspace{3mm}r\ge r_1 \ , \nonumber\\
J(r)&=&J_0e^{\alpha (r-r_0)},\hspace{3mm}r\le r_0 \ . \label{jl}
\end{eqnarray}
The rise at short distances together with the decay at large distances result in a peak at intermediate distances in agreement with observations \cite{86SP}. In the intermediate range the exchange integral can be fitted by a polynomial to ensure continuity and
differentiability.

\begin{figure}[!h]
\centering
\epsfig{width=.38\textwidth,file=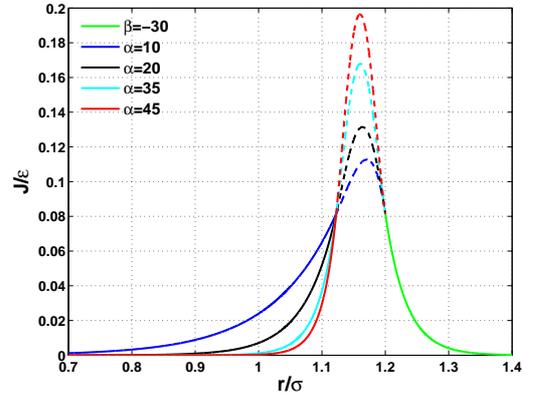}
\caption{The exchange integral computed by fitting to experimental data in Nickel. The interparticle
distance is measured in units of $\sigma$ which is the length appearing in the mechanical two-body interaction.
The parameters $\alpha$ and $\beta$ were defined in Eq. (A2).}
\label{fig1}
\end{figure}

An example of the exchange integral function for Nickel based on the Monte Carlo / Molecular Dynamics
results of \cite{GR96} is shown in Fig. \ref{fig1}. Note that $J(r)$ is not a monotonically
decreasing function of $r$. In fact,
%%%%%%%%%%%%%%%%%%%%%%%%%%
%%%%%%%%%%%%%%%%%%%%%%%%%%
the overlap integral tends to peak roughly at the typical distance between particles,
set by the mechanical interaction (\ref{umech}), denoted as $\sigma$ in the ordinate of Fig. \ref{fig1}.
Thus the effects of strain and magnetic fields will be crucially dependent on where exactly $J(r)$ peaks
with respect to $\sigma$. For example the present form of $J(r)$ means that for crystalline Nickel
the magnetization should decrease under increasing pressure.

\section{Appendix B: Calculation of Coefficients}
The coefficients in Eqs. \ref{eigenvalueDerivative} are determined by

\begin{widetext}
\begin{eqnarray}
\label{cont}
a^{(\gamma )}_k  &=&  \Xv^{(\gamma )} \cdot \eigK{k}\ , \quad
c^{(\gamma )}_{k\ell}  =  \frac{\partial \Hes}{\partial \gamma} : \eigK{k}\eigK{\ell}\ , \quad
a^{(b)}_k  =  \Xv^{(b)} \cdot \eigK{k}\ ,
c^{(b)}_{k\ell}  =  \frac{\partial \Hes}{\partial B} : \eigK{k}\eigK{\ell}\ ,
b^{(r)}_{k\ell m}  =  \Tesr\  \vdots\  \eigK{k} \eigK{\ell} \eigK{m}\ , \nonumber \\
b^{(\phi )}_{k\ell m} & = & \Tess\  \vdots\  \eigK{k} \eigK{\ell} \eigK{m}\ , \label{contractions}
\end{eqnarray}
\end{widetext}
where $\eigK{k}$ is the eigenfunction of the Hessian associated with eigenvalue $\lambda_k$,
\begin{equation}
\label{xia}
\Xv^{(\gamma )} =
\begin{pmatrix}
  \Xv^{(\gamma, r )}\\
  \Xv^{(\gamma, \phi)}
\end{pmatrix} \ , \quad
\Xv^{(b)} =
\begin{pmatrix}
  \Xv^{(b, r )}\\
  \Xv^{(b, \phi )}
\end{pmatrix}
\end{equation}
\begin{eqnarray}
\label{mismatch}
\Xv^{(\gamma, r )}  &\equiv&  \frac{\partial^2U}{\partial \gamma \partial \rv},\quad
\Xv^{(\gamma, \phi )}  \equiv   \frac{\partial^2U}{\partial \gamma \partial \phi} \nonumber\\
\Xv^{(b, r)}  &\equiv &  \frac{\partial^2U}{\partial B \partial \rv} \ , \quad
\Xv^{(b, \phi )}  \equiv   \frac{\partial^2U}{\partial B \partial \phi} .
\end{eqnarray}
Finally,
\begin{equation}
\Tesr =
\begin{pmatrix}
  \Trrr & \Trsr\\
  \Tsrr & \Tssr
\end{pmatrix}\ , 
\Tess =
\begin{pmatrix}
  \Trrs & \Trss\\
  \Tsrs & \Tsss
\end{pmatrix}
\end{equation}
and we have denoted the third-order tensor of potential energy derivatives as, for example,
$\Trrr_{ijk} \equiv \frac{\partial^3U}{\partial \rv_k\partial \rv_j\partial \rv_i}$ and $\Tssr_{ijk} \equiv \frac{\partial^3U}{\partial \phi_k\partial \phi_j\partial \rv_i}$.

\end{document}